\documentclass[pra,twocolumn,amsmath,amssymb]{revtex4-1}
\usepackage{graphicx}
\usepackage{amsmath}
\usepackage{MnSymbol}
\usepackage{mathrsfs}
\usepackage{colortbl}
\usepackage[table]{xcolor}
\usepackage{tabularx}
\makeatletter

\newenvironment{FIG.here}
  {\def\@captype{FIG.}}
  {}
\makeatother

\begin{document}


\title{Relativistic properties of a molecule: energy, linear momentum, angular momentum and boost momentum to order $1/c^2$}

\author{Robert P. Cameron}
\email{robert.p.cameron@strath.ac.uk, www.ytilarihc.com}
\address{SUPA and Department of Physics, University of Strathclyde, Glasgow G4 0NG, United Kingdom}

\author{J. P. Cotter}
\address{Centre for Cold Matter, Blackett Laboratory, Imperial College London, Prince Consort Road, London SW7 2AZ, United Kingdom}

\begin{abstract}
We give an explicit and general description of the energy, linear momentum, angular momentum and boost momentum of a molecule to order $1/c^2$, where it necessary to take account of kinetic contributions made by the electrons and nuclei as well as electromagnetic contributions made by the intramolecular field. A wealth of interesting subtleties are encountered that are not seen at order $1/c^0$, including relativistic Hall shifts, anomalous velocities and hidden momenta. Some of these have well known analogues in solid state physics.
\end{abstract}

\date{\today}
\maketitle


\section{Introduction}
\label{Introduction}
Energy, linear momentum, angular momentum and boost momentum are of fundamental importance, being associated with the Poincar\'{e} group: translations in time, translations in space \footnote{More precisely, translations with respect to \textit{Cartesian} spatial coordinates: it has recently been shown that translational symmetry with respect to other spatial coordinates can give rise to different associations and that this sheds light upon the distinction between canonical and symmetric energy-momentum tensors \cite{Cameron15a}.}, circular rotations in space and hyperbolic rotations in spacetime (Lorentz transformations or boosts) \cite{Landau, Bessel-Hagen21a, Griffiths99a, Jackson99a, Barnett11a, Cameron12a, Bliokh13a, Bliokh18a}. At a fundamental level, the conservation of these quantities \cite{Noether18a, Bessel-Hagen21a} facilitates the examination and exploitation of molecules in the laboratory, through changes brought about by interactions: frequency spectra yield knowledge about molecular energy and angular momentum differences, deflection techniques utilise modifications of molecular linear momentum and boost momentum and so on.

The present authors recently began a high-precision molecular calculation and, to our surprise, were unable to find a complete discussion of the aforementioned quantities for a molecule beyond that afforded by the `$\sum p^2/2m+V$' Hamiltonian, which is valid to order $1/c^0$ only. This shortfall has encouraged us to write the present paper, in which we give an explicit and general description of the energy, linear momentum, angular momentum and boost momentum of a molecule to order $1/c^2$. Our paper should prove increasingly useful as molecules are examined and exploited with increasing precision \cite{Darquie10a, Hudson11a, Pavenello12a, Cameron14a, Jusko16a, Biesheuvel16a, Cameron16a, Asselin17a, Baron17a, Gregory17a, Truppe17a, Cairncross17a, Beyer18a}. Our approach is general and our results are of relevance to any molecule (or atom). They might prove especially useful for molecules containing heavy atoms, where relativistic effects can be pronounced.

We consider ourselves to be in an inertial frame of reference with right-handed Cartesian coordinates $x$, $y$ and $z$ and time $t$. The SI system of units is adopted and the Einstein summation convention \cite{Einstein16a} is to be understood, with subscripts $a$, $b$, $c$, $\dots$ running over $x$, $y$ and $z$. We adopt the Coulomb gauge \cite{Maxwell61a} and work in the Schr\"{o}dinger picture of time dependence.


\section{Molecular model and Hamiltonian}
\label{Molecule}
In the present section, we define our chosen molecular model and Hamiltonian.

\subsection{Molecular model}
We consider a single molecule in isolation: an electrically neutral collection of electrons (subscript $i$) and nuclei (subscript $j$), bound together by electromagnetic interactions in the absence of external influences. We refer to the electrons and nuclei collectively as `the particles' (subscript $k$). Our molecular Hamiltonian $\hat{H}$ (introduced below) is effectively truncated at order $1/c^2$ and we therefore neglect terms of order $1/c^3$ or smaller.

\subsubsection{Electrons}
We treat the $i$th electron as a point particle of rest mass $m_i=m_e$, mean position $\hat{\mathbf{r}}_i=\mathbf{r}_i$ \cite{Foldy50a}, canonical linear momentum $\hat{\mathbf{p}}_i=-\textrm{i}\hbar\pmb{\nabla}_i$, electric charge $q_i=-e$ and magnetic-dipole moment $\hat{\mathbf{m}}_i=\gamma_i \hat{\mathbf{s}}_i$, with $\pmb{\nabla}_i$ the gradient with respect to $\mathbf{r}_i$, $\gamma_i=-e/m_e$ the gyromagnetic ratio and $\hat{\mathbf{s}}_i=\hbar\hat{\pmb{\sigma}}_i/2$ the mean spin \cite{Foldy50a}, where $\hat{\pmb{\sigma}}_i$ is a pseudovector of Pauli matrices \cite{Pauli27a, Dirac28a}. We identify an (effective) finite-size parameter $R_i$, defined such that $q_i R_i^2=-3e\lambdabar_e^2/4$ \cite{Itoh65a}, with $\lambdabar_e=\hbar/m_e c$ the reduced Compton wavelength \cite{Compton23a} of the electron: $R_i$ accounts for a smearing due to the electron's \textit{Zitterbewegung} \cite{Schrodinger30a,Schrodinger31a}. 

\subsubsection{Nuclei}
For the $j$th nucleus, we adopt one of the following three treatments, depending upon the value of the spin quantum number $s_j$. 
\begin{itemize}
\item For $s_j=0$, we model the nucleus as a point particle of rest mass $m_j$, mean position $\hat{\mathbf{r}}_j=\mathbf{r}_j$, canonical linear momentum $\hat{\mathbf{p}}_j=-\textrm{i}\hbar\pmb{\nabla}_j$ and electric charge $q_j=Z_je$, with $\pmb{\nabla}_j$ the gradient with respect to $\mathbf{r}_j$ and $Z_j$ the atomic number.  \\
\item For $s_j=1/2$, we extend the $s_j=0$ model by granting the nucleus a non-vanishing magnetic-dipole moment $\hat{\mathbf{m}}_j=\gamma_j \hat{\mathbf{s}}_j$, with $\gamma_j$ the gyromagnetic ratio and $\hat{\mathbf{s}}_j=\hbar\hat{\pmb{\sigma}}_j/2$ the mean spin, where $\hat{\pmb{\sigma}}_j$ is a pseudovector of Pauli matrices. \\
\item For $s_j\in\{1, 3/2, \dots\}$, we extend the $s_j=1/2$ model by modifying the explicit form of the spin to $\hat{\mathbf{s}}_j=\hbar s_j\hat{\pmb{\lambda}}_j$ whilst granting the nucleus a non-vanishing electric-quadrupole moment $\hat{\Theta}_{jab}=eQ_j [3(\hat{s}_{ja}\hat{s}_{jb}+\hat{s}_{jb}\hat{s}_{ja})/2-\hbar^2\delta_{ab} s_j (s_j+1) ]/2\hbar^2 s_j(2s_j-1)$, with $\hat{\pmb{\lambda}}_j$ a pseudovector of the relevant higher-order spin matrices and $Q_j$ the electric-quadrupole-moment parameter \cite{Casimir36a,Ramsey56a}. 
\end{itemize}
We account for the finite-size of the $j$th nucleus further via the radius parameter $R_j$ \cite{Helgaker} and treat the $\hat{\Theta}_{jab}$ and the $R_j^2$ as if they are of order $1/c^2$, as their associated contributions are `small'. $R_j$ might include contributions due to nuclear \textit{Zitterbewegung} \cite{Pachucki95a}. It is to be understood that $\hat{\mathbf{m}}_j/\gamma_j=\hat{\mathbf{s}}_j=0$ for $s_j=0$.

\subsection{The molecular Hamiltonian}
We take the Hamiltonian that governs our molecule to be \cite{Uhlenbeck26a, Thomas26a, Heisenberg26a, Thomas27a, Darwin28a, Gaunt29a, Breit29a, Fermi30a, Breit30a, Breit32a, Casimir36a, Kellogg39a, Henderson48a, Chraplyvy53a, Chraplyvy53b, Gunther-Mohr54a, Itoh65a, Helgaker}
\begin{eqnarray}
\hat{H}&=&c^2\sum_k m_k+\frac{1}{2}\sum_k\frac{\hat{p}_k^2}{m_k}+\frac{1}{2}\sum_k q_k \hat{\Phi}_k \nonumber \\
&-&\frac{1}{8c^2}\sum_k\frac{\hat{p}_k^4}{m_k^3}+\frac{1}{12}\sum_kq_kR_k^2\nabla_k^2\hat{\Phi}_k^q \nonumber \\
&-&\frac{1}{c^2}\sum_k\left(1-\frac{q_k}{2m_k\gamma_k}\right)\hat{\mathbf{m}}_k\cdot \left( \frac{\hat{\mathbf{p}}_k}{m_k} \times \pmb{\nabla}_k \hat{\Phi}_k^q\right) \nonumber \\
&-&\frac{1}{2}\sum_k \frac{q_k}{m_k}\hat{\mathbf{p}}_k\cdot\hat{\mathbf{A}}_k-\frac{1}{2}\sum_k \hat{\mathbf{m}}_k\cdot(\pmb{\nabla}_k\times\hat{\mathbf{A}}_k) \nonumber \\
&+&\frac{1}{6}\sum_j\hat{\Theta}_{jab}\partial_{ja}\partial_{jb}\hat{\Phi}_j^q, \label{condensed form} 
\end{eqnarray} 
with
\begin{eqnarray}
\hat{\Phi}_k&=&\hat{\Phi}_k^q+\hat{\Phi}_k^R+\hat{\Phi}_k^\Theta \\ 
\hat{\mathbf{A}}_k&=&\hat{\mathbf{A}}_k^\mathbf{m}+\hat{\mathbf{A}}_k^\mathbf{v}
\end{eqnarray}
the intramolecular scalar and magnetic vector potentials seen by the $k$th particle at its mean position $\hat{\mathbf{r}}_k$.
\begin{eqnarray}
\hat{\Phi}_k^q&=&\frac{1}{4\pi\epsilon_0}\sum_{k'\ne k}\frac{q_{k'}}{\hat{r}_{kk'}}, \\
\hat{\Phi}_k^R&=&-\frac{1}{6\epsilon_0}\sum_{k'\ne k}q_{k'} R_{k'}^2\delta^3(\hat{\mathbf{r}}_{kk'}) \\
\hat{\Phi}_k^\Theta&=&\frac{1}{12\pi\epsilon_0}\sum_{j\ne k}\frac{\hat{\Theta}_{jab}}{\hat{r}_{kj}^3}\left(\frac{3\hat{r}_{kja}\hat{r}_{kjb}}{\hat{r}_{kj}^2}-\delta_{ab}\right)
\end{eqnarray}
account for the electric charges of the other particles, the finite sizes of the other particles and the electric-quadrupole moments of the other nuclei.
\begin{eqnarray}
\hat{\mathbf{A}}_k^\mathbf{m}&=&\frac{\mu_0}{4\pi}\sum_{k'\ne k}\frac{\hat{\mathbf{m}}_{k'}\times\hat{\mathbf{r}}_{kk'}}{\hat{r}_{kk'}^3} \\
\hat{\mathbf{A}}_k^\mathbf{v}&=&\frac{\mu_0}{16\pi}\sum_{k'\ne k}\frac{q_{k'}}{m_{k'}}\Bigg[\frac{1}{\hat{r}_{kk'}}\hat{\mathbf{p}}_{k'}+\hat{\mathbf{p}}_{k'}\frac{1}{\hat{r}_{kk'}} \nonumber \\
&+&\hat{\mathbf{r}}_{kk'}\frac{1}{\hat{r}_{kk'}^3}(\hat{\mathbf{r}}_{kk'}\cdot\hat{\mathbf{p}}_{k'})+(\hat{\mathbf{p}}_{k'}\cdot\hat{\mathbf{r}}_{kk'})\frac{1}{\hat{r}_{kk'}^3}\hat{\mathbf{r}}_{kk'}\Bigg]
\end{eqnarray}
account for the intrinsic magnetic moments and orbital motions of the other particles. 

The molecular Hamiltonian $\hat{H}$ can be obtained as the non-relativistic limit \cite{Chraplyvy53a, Chraplyvy53b} of the Breit Hamiltonian \cite{Breit29a, Breit30a, Breit32a} or from quantum electrodynamics \cite{Itoh65a}, with nuclear contributions added heuristically. At the time of writing, there is ``no unambiguous and rigorous procedure for constructing molecular Hamiltonians for particles other than electrons'' \cite{Helgaker}. The difficulty stems in part from the composite nature of nucleons and nuclei, as well as the absence of universally accepted field equations for higher-spin particles: the Rarita-Schwinger field equation for spin-$3/2$ particles \cite{Rarita41}, for example, has been accused of acausal behaviour \cite{Velo69a}. The various terms that comprise $\hat{H}$ are described in more detail in the appendix.

The list of intramolecular interactions included in the molecular Hamiltonian $\hat{H}$ is almost exhaustive to order $1/c^2$. One might add additional terms due to the internal structures of the nuclei \cite{Bohr50a, Ramsey56a} or non-electromagnetic corrections due to the parity-violating influence of the weak interaction \cite{Lee56a, Wu57a, Rein74a, Darquie10a, Barron04a, Cohen11a,Asselin17a}, for example. In the present paper we do not account explicitly for vacuum effects, but it might prove interesting to do so \cite{Manjavacas10a,Sonnleitner17a}.


\section{Particle positions, velocities and linear momenta}
\label{Particle velocities}
In the present section, we focus our attention upon the $k$th particle and highlight some subtleties inherent to the description of its postion, velocity and linear momentum. An understanding of these subtleties is important for the next section, where we consider our molecule in its entirety.

\subsection{Positions}
In spite of its name (and simple operator representative), the mean position $\hat{\mathbf{r}}_k$($=\mathbf{r}_k$) does \textit{not} embody the average position of an electron or spinning nucleus: the `mean' terminology introduced in \cite{Foldy50a} for $\hat{\mathbf{r}}_k$ and other quantities such as the mean spin $\hat{\mathbf{s}}_k$ is something of a misnomer \cite{Barut81a}. The role is more closely filled by the kinetic position $\hat{\bar{\mathbf{r}}}_k$ \cite{Born35a}, which differs from $\hat{\mathbf{r}}_k$ by the position difference \cite{Pryce35a}
\begin{eqnarray}
\hat{\pmb{\delta}}_k&=&\hat{\bar{\mathbf{r}}}_k-\hat{\mathbf{r}}_k \nonumber \\
&=&\frac{\hat{\mathbf{p}}_k\times\hat{\mathbf{s}}_k}{2c^2m_k^2}.
\end{eqnarray}
To appreciate the distinction between $\hat{\mathbf{r}}_k$ and $\hat{\bar{\mathbf{r}}}_k$ in simple terms, consider a translating spinning wheel: $\hat{\mathbf{r}}_k$ is analogous to the axle whereas $\hat{\bar{\mathbf{r}}}_k$ is analogous to the element-weighted centre of the wheel, which is shifted away from the axle as different elements on the rim have different speeds and are Lorentz contracted by different amounts. $\hat{\pmb{\delta}}_k$ might thus be regarded as a manifestation of the relativistic Hall effect \cite{Bliokh12a} and is intimately associated with Thomas precession \cite{Muller92a}. For a spinless nucleus, $\hat{\pmb{\delta}}_j=0$. Also of importance is the position 
\begin{equation}
\hat{\bar{\mathbf{r}}}_k'=\hat{\mathbf{r}}_k+\left(\frac{2m_k\gamma_k}{q_k}-1\right)\hat{\pmb{\delta}}_k
\end{equation}
of the centre of electric charge. For an electron, $\hat{\bar{\mathbf{r}}}_i'$ coincides with $\hat{\bar{\mathbf{r}}}_i$ \cite{Bliokh17a} and, indeed, the position about which the electron exhibits its \textit{Zitterbewegung} \cite{Barut81a}. For a spinning nucleus, $\hat{\bar{\mathbf{r}}}_j'$ is distinct from $\hat{\bar{\mathbf{r}}}_j$. Note that the distinctions between $\hat{\mathbf{r}}_k$, $\hat{\bar{\mathbf{r}}}_k$ and $\hat{\bar{\mathbf{r}}}_k'$ first become apparent at order $1/c^2$ and so do not enter into discussions founded upon the `$\sum p^2/2m+V$' molecular Hamiltonian, which is valid to order $1/c^0$ only. 

Position differences like $\hat{\pmb{\delta}}_k$ are well known for electrons in the solid state and can be regarded as Berry connections in momentum space \cite{Adams59a,
Bliokh05a,Berard06a,Chang08a,Xiao10a,Bliokh11a,Takahashi15a, Bliokh17a}.

\subsection{Velocities}
When the $k$th particle is considered in isolation, the position difference $\hat{\pmb{\delta}}_k$ is found to be independent of time. The mean position $\hat{\mathbf{r}}_{k}$, kinetic position $\hat{\bar{\mathbf{r}}}_k$ and position $\hat{\bar{\mathbf{r}}}_k'$ of the centre of electric charge then differ by constant amounts. When the particle is considered as a constituent of our molecule, however, $\hat{\pmb{\delta}}_k$ can vary in time, as the intramolecular electromagnetic field can vary $\hat{\mathbf{p}}_k\times\hat{\mathbf{s}}_k$:
\begin{equation}
\frac{\textrm{d}\hat{\pmb{\delta}}_k}{\textrm{d}t}=-\frac{q_k\hat{\mathbf{s}}_k\times\pmb{\nabla}_k\hat{\Phi}_k^q}{2c^2m_k^2}+\textrm{O}(\tfrac{1}{c^3}), \label{deltadot}
\end{equation}
where we have made use of $\textrm{d}\hat{\mathbf{p}}_k/\textrm{d}t=-q_k\pmb{\nabla}_k\hat{\Phi}_k^q+\textrm{O}(1/c^1)$ and $\textrm{d}\hat{\mathbf{s}}_k/\textrm{d}t=0+\textrm{O}(1/c^1)$. It follows that the mean velocity
\begin{eqnarray}
\hat{\mathbf{v}}_k&=&\frac{\textrm{d}\hat{\mathbf{r}}_k}{\textrm{d}t} \nonumber \\
&=&\frac{\hat{\mathbf{p}}_k}{m_k}-\frac{\hat{p}_k^2\hat{\mathbf{p}}_k}{2c^2 m_k^3}-\frac{q_k\hat{\mathbf{s}}_k\times\pmb{\nabla}_k\hat{\Phi}_k^q}{2c^2m_k^2}+\frac{\hat{\mathbf{m}}_k\times\pmb{\nabla}_k\hat{\Phi}_k^q}{c^2m_k}-\frac{q_k\hat{\mathbf{A}}_k}{m_k} \nonumber \\
&& \label{meanvelocity}
\end{eqnarray}
does not necessarily coincide with the kinetic velocity
\begin{eqnarray}
\hat{\bar{\mathbf{v}}}_k&=&\frac{\textrm{d}\hat{\bar{\mathbf{r}}}_k}{\textrm{d}t} \nonumber \\
&=&\frac{\hat{\mathbf{p}}_k}{m_k}-\frac{\hat{p}_k^2\hat{\mathbf{p}}_k}{2c^2 m_k^3}+\frac{\hat{\mathbf{m}}_k\times\pmb{\nabla}_k\hat{\Phi}_k^q}{c^2m_k}-\frac{q_k\hat{\mathbf{A}}_k}{m_k}+\textrm{O}(\tfrac{1}{c^3}), \nonumber \\
&& \label{kineticvelocity}
\end{eqnarray}
or the velocity 
\begin{eqnarray}
\hat{\bar{\mathbf{v}}}_k'&=&\frac{\textrm{d}\hat{\bar{\mathbf{r}}}_k'}{\textrm{d}t} \nonumber \\
&=&\hat{\mathbf{v}}_k+\left(\frac{2m_k\gamma_k}{q_k}-1\right)\frac{\textrm{d}\hat{\pmb{\delta}}_k}{\textrm{d}t} \label{chargevelocity}
\end{eqnarray}
of the centre of charge.

Velocity contributions like the $\hat{\mathbf{s}}_k$- and $\hat{\mathbf{m}}_k$-dependent terms seen in (\ref{meanvelocity}) and (\ref{kineticvelocity}) are also known for electrons in the solid state and are sometimes referred to as being `anomalous' \cite{Adams59a,Bliokh05a,Berard06a,Chang08a,Xiao10a,Takahashi15a}.

\subsection{Linear momenta}
The total kinetic linear momentum $\hat{\pmb{\pi}}_k$ follows from the canonical linear momentum $\hat{\mathbf{p}}_k$ as $\hat{\pmb{\pi}}_k=\hat{\mathbf{p}}_k-q_k\hat{\mathbf{A}}_k$. A revealing expression for $\hat{\mathbf{p}}_k$ can be obtained by rearranging (\ref{kineticvelocity}) and interating:
\begin{eqnarray}
\hat{\mathbf{p}}_k&=&m_k\hat{\bar{\mathbf{v}}}_k+\frac{\hat{p}_k^2\hat{\mathbf{p}}_k}{2c^2 m_k^2}+q_k\hat{\mathbf{A}}_k-\frac{\hat{\mathbf{m}}_k\times\pmb{\nabla}_k\hat{\Phi}_k^q}{c^2}+\textrm{O}(\tfrac{1}{c^3}) \nonumber \\
&=&m_k\hat{\bar{\mathbf{v}}}_k+\frac{ \hat{\pi}_k^2\hat{\bar{\mathbf{v}}}_k+\hat{\bar{\mathbf{v}}}_k\hat{\pi}_k^2}{4c^2m_k}+q_k\hat{\mathbf{A}}_k-\frac{\hat{\mathbf{m}}_k\times\pmb{\nabla}_k\hat{\Phi}_k^q}{c^2} \nonumber \\
&+&\textrm{O}(\tfrac{1}{c^3}). \label{Particlecanonicalp}
\end{eqnarray}
Evidently, $\hat{\mathbf{p}}_k$ is comprised of relativistically corrected kinetic linear momentum terms ($m_k\hat{\bar{\mathbf{v}}}_k+ (\hat{\pi}_k^2\hat{\bar{\mathbf{v}}}_k+\hat{\bar{\mathbf{v}}}_k\hat{\pi}_k^2)/4c^2m_k$), an electromagnetic linear momentum term ($q_k\hat{\mathbf{A}}_k$) and a hidden momentum term ($-\hat{\mathbf{m}}_k\times\pmb{\nabla}_k\hat{\Phi}_k^q/c^2$) with the prototypical form \cite{Shockley67a, vanVleck69a, Griffiths99a, Babson09a, Cameron18a}. For an electron, we attribute this hidden momentum to a modification of the electron's \textit{Zitterbewegung} by the intramolecular Coulomb electric field $-\pmb{\nabla}_k\hat{\Phi}_k^q$. For more details, see our recent paper on hidden momentum \cite{Cameron18a}. Note that the distinction between $\hat{\pmb{\pi}}_k$ and $\hat{\mathbf{p}}_k$ first becomes apparent at order $1/c^2$; that
it is necessary to identify the kinetic velocity $\hat{\bar{\mathbf{v}}}_k$ in the first term of (\ref{Particlecanonicalp}), rather than the mean velocity $\hat{\mathbf{v}}_k$, for example, and that $\hat{\pmb{\pi}}_k$ contains within it the hidden momentum $-\hat{\mathbf{m}}_k\times\pmb{\nabla}_k\hat{\Phi}_k^q/c^2$.


\section{Main results}
In the present section, we give our main results: an explicit and general description of the energy, linear momentum, angular momentum and boost momentum of a molecule to order $1/c^2$. Before describing each of these properties for our molecule, we will consider the analogous results for a simpler system that is well understood: a collection of structureless point particles interacting electromagnetically in the classical domain \cite{Landau, Griffiths99a, Jackson99a}. We use tildes to distinguish this system from the molecule. 

The particles give rise to electric charge and current densities
\begin{eqnarray}
\tilde{\rho}&=&\sum_k\tilde{q}_k\delta^3\left(\mathbf{r}-\tilde{\mathbf{r}}_k\right) \label{chargedensitydefinition} \\
\tilde{\mathbf{j}}&=&\sum_k\tilde{q}_k\tilde{\mathbf{v}}_k\delta^3\left(\mathbf{r}-\tilde{\mathbf{r}}_k\right), \label{currentdensitydefinition}
\end{eqnarray}
with $\tilde{q}_k$ the electric charge of the $k$th particle, $\tilde{\mathbf{r}}_k$ the position of the $k$th particle and $\tilde{\mathbf{v}}_k=\textrm{d}\tilde{\mathbf{r}}_k/\textrm{d}t$ the velocity of the $k$th particle. The trajectory of the $k$th particle is governed by the Lorentz force law \cite{Lorentz95a}
\begin{equation} 
\frac{\textrm{d}\tilde{\pmb{\pi}}_k}{\textrm{d}t}=\tilde{q}_k\left[\mathbf{E}\left(\tilde{\mathbf{r}}_k\right)+\tilde{\mathbf{v}}_k\times\mathbf{B}\left(\tilde{\mathbf{r}}_k\right)\right], \label{Lorentzforce}
\end{equation}
with
\begin{equation}
\tilde{\pmb{\pi}}_k=\frac{\tilde{m}_k\tilde{\mathbf{v}}_k}{\sqrt{1-\tilde{v}_k^2/c^2}}
\end{equation}
the kinetic momentum of the $k$th particle. The electric and magnetic fields
\begin{eqnarray}
\tilde{\mathbf{E}}&=&-\pmb{\nabla}\tilde{\Phi}-\frac{\partial\tilde{\mathbf{A}}}{\partial t} \\
\tilde{\mathbf{B}}&=&\pmb{\nabla}\times\tilde{\mathbf{A}}
\end{eqnarray}
are governed by Maxwell's equations \cite{Maxwell61a}
\begin{eqnarray}
\pmb{\nabla}\cdot\tilde{\mathbf{E}}&=&\frac{\tilde{\rho}}{\epsilon_0}, \label{AGauss1} \\
\pmb{\nabla}\cdot\tilde{\mathbf{B}}&=&0, \label{AGauss2} \\
\pmb{\nabla}\times\tilde{\mathbf{E}}&=&-\frac{\partial\tilde{\mathbf{B}}}{\partial t} \label{AFaraday-Lenz} \\
\pmb{\nabla}\times\tilde{\mathbf{B}}&=&\epsilon_0\tilde{\mathbf{j}}+\epsilon_0\mu_0\frac{\partial\tilde{\mathbf{E}}}{\partial t}, \label{AAmpere-Maxwell}
\end{eqnarray}
with $\tilde{\Phi}$ and $\tilde{\mathbf{A}}$ the scalar and magnetic-vector potentials. We treat $\tilde{\mathbf{A}}$ as if its largest contribution is of order $1/c^2$.


\subsection{Energy}
\label{Energy}
\subsubsection{Classical system}
The total energy of a collection of structureless point particles interacting electromagnetically in the classical domain is \cite{Griffiths99a, Jackson99a}
\begin{equation}
\tilde{W}=\tilde{W}_{\textrm{rest+kin}}+\tilde{W}_\textrm{em},
\end{equation}
with
\begin{eqnarray}
\tilde{W}_{\textrm{rest+kin}}&=&\sum_k\sqrt{c^4\tilde{m}_k^2+c^2\tilde{\pi}_k^2} \nonumber \\
&=&c^2\sum_k\tilde{m}_k+\frac{1}{2}\sum_k\frac{\tilde{\pi}_k^2}{\tilde{m}_k}-\frac{1}{8c^2}\sum_k\frac{\tilde{\pi}_k^4}{\tilde{m}_k^2}+\textrm{O}(\tfrac{1}{c^3}) \nonumber \\
&& \label{Wmech}
\end{eqnarray}
the total rest and kinetic energy of the particles and
\begin{eqnarray}
\tilde{W}_\textrm{em}&=&\frac{\epsilon_0}{2}\int\left(\tilde{E}^2+c^2\tilde{B}^2\right)\textrm{d}^3\mathbf{r} \nonumber \\
&=&\frac{1}{2}\int\tilde{\rho}\tilde{\Phi}\textrm{d}^3\mathbf{r}+\frac{1}{2}\int\tilde{\mathbf{j}}\cdot\tilde{\mathbf{A}}\textrm{d}^3\mathbf{r}+\textrm{O}(\tfrac{1}{c^3}), \nonumber \\
&=&\frac{1}{2}\sum_k\tilde{q}_k\tilde{\Phi}(\tilde{\mathbf{r}}_k)+\frac{1}{2}\sum_k\tilde{q}_k\tilde{\mathbf{v}}_k\cdot\tilde{\mathbf{A}}(\tilde{\mathbf{r}}_k)+\textrm{O}(\tfrac{1}{c^3}) \nonumber \\
\label{Wfield}
\end{eqnarray}
the total energy of the electromagnetic field.

\subsubsection{Molecule}
We identify the total energy of our molecule as being
\begin{equation}
\hat{W}=\hat{H}.
\end{equation}
$\hat{W}$ generates translations in time:
\begin{eqnarray}
\textrm{e}^{\textrm{i}\tau\hat{W}/\hbar}\hat{\mathbf{s}}_k\textrm{e}^{-\textrm{i}\tau\hat{W}/\hbar}&=&\hat{\mathbf{s}}_k+\tau\frac{\textrm{d}\hat{\mathbf{s}}_k}{\textrm{d}t}+\textrm{O}(\tau^2), \label{Wtranslate1} \\
\textrm{e}^{\textrm{i}\tau\hat{W}/\hbar}\hat{\mathbf{r}}_k\textrm{e}^{-\textrm{i}\tau\hat{W}/\hbar}&=&\hat{\mathbf{r}}_k+\tau\frac{\textrm{d}\hat{\mathbf{r}}_k}{\textrm{d}t}+\textrm{O}(\tau^2) \\
\textrm{e}^{\textrm{i}\tau\hat{W}/\hbar}\hat{\mathbf{p}}_k\textrm{e}^{-\textrm{i}\tau\hat{W}/\hbar}&=&\hat{\mathbf{p}}_k+\tau\frac{\textrm{d}\hat{\mathbf{p}}_k}{\textrm{d}t}+\textrm{O}(\tau^2). \label{Wtranslate3}
\end{eqnarray}
The conservation law
\begin{equation}
\frac{\textrm{d}\hat{W}}{\textrm{d}t}=0 \label{energyconservation}
\end{equation}
shows that this is a symmetry transformation for the molecule \cite{Noether18a}, embodying invariance under translations in time.

It is not immediately obvious that the energy $\hat{W}$ of the molecule has a form analogous to the energy $\tilde{W}$ of the classical system described above. To proceed, we observe that (\ref{condensed form}) can be rewritten in the form
\begin{eqnarray}
\hat{W}&=&\sum_k m_kc^2+\frac{1}{2}\sum_k\frac{\hat{p}_k^2}{m_k}-\sum_k \frac{q_k}{m_k}\hat{\mathbf{p}}_k\cdot\hat{\mathbf{A}}_k-\frac{1}{8c^2}\sum_k\frac{\hat{p}_k^4}{m_k^3} \nonumber \\
&+&\frac{1}{2}\sum_k q_k \hat{\Phi}_k^q-\frac{1}{c^2}\sum_k\left(1-\frac{q_k}{2m_k\gamma_k}\right)\hat{\mathbf{m}}_k\cdot \left( \frac{\hat{\mathbf{p}}_k}{m_k} \times \pmb{\nabla}_k \hat{\Phi}_k^q\right) \nonumber \\
&+&\frac{1}{2}\sum_k \frac{q_k}{m_k}\hat{\mathbf{p}}_k\cdot\hat{\mathbf{A}}_k \nonumber \\
&+&\frac{1}{6}\sum_kq_kR_k^2\nabla_k^2\hat{\Phi}_k^q-\frac{1}{2}\sum_k \hat{\mathbf{m}}_k\cdot(\pmb{\nabla}_k\times\hat{\mathbf{A}}_k) \nonumber \\
&+&\frac{1}{3}\sum_j\hat{\Theta}_{jab}\partial_{ja}\partial_{jb}\hat{\Phi}_j^q, \label{W0explicit}
\end{eqnarray} 
where we have made use of the following equalities:
\begin{eqnarray}
\frac{1}{2}\sum_kq_k\hat{\Phi}_k^R&=&\frac{1}{12}\sum_kq_kR_k^2\nabla_k^2\hat{\Phi}_k^q \\
\frac{1}{2}\sum_kq_k\hat{\Phi}_k^\Theta&=&\frac{1}{6}\sum_j\hat{\Theta}_{jab}\partial_{jab}\partial_{ja}\partial_{jb}\hat{\Phi}_j^q.
\end{eqnarray}
The first line on the right-hand side of (\ref{W0explicit}) can be recast as
\begin{eqnarray}
&&\sum_km_kc^2+\frac{1}{2}\sum_k\frac{\hat{p}_k^2}{m_k}-\sum_k\frac{q_k}{m_k}\hat{\mathbf{p}}_k\cdot\hat{\mathbf{A}}_k-\frac{1}{8c^2}\sum_k\frac{\hat{p}_k^4}{m_k^3} \nonumber \\
&=&\sum_k m_k c^2+\frac{1}{2}\sum_k\frac{\hat{\pi}_k^2}{m_k}-\frac{1}{8c^2}\sum_k\frac{\hat{\pi}_k^4}{m_k^3}+\textrm{O}(\tfrac{1}{c^3}).
\end{eqnarray}
The second line can be recast as
\begin{eqnarray}
&&\frac{1}{2}\sum_k q_k \hat{\Phi}_k^q-\frac{1}{c^2}\sum_k\left(1-\frac{q_k}{2m_k\gamma_k}\right)\hat{\mathbf{m}}_k\cdot \left( \frac{\hat{\mathbf{p}}_k}{m_k} \times \pmb{\nabla}_k \hat{\Phi}_k^q\right) \nonumber \\
&=&\frac{1}{2}\sum_kq_k\hat{\Phi}_k^q+\sum_kq_k\left(\hat{\bar{\mathbf{r}}}_k'-\hat{\mathbf{r}}_k\right)\cdot\pmb{\nabla}_k\hat{\Phi}_k^q, \label{Taylor}
\end{eqnarray}
which is in accord with the idea that $\hat{\bar{\mathbf{r}}}'_k$ is the position of the centre of charge of the $k$th particle: one can interpret (\ref{Taylor}) as a Taylor series expansion of the expected electrostatic interaction expressed in terms of the $\hat{\bar{\mathbf{r}}}'_k$, with the expansion taken about the mean positions $\hat{\mathbf{r}}_k$ and truncated at order $1/c^2$. An alternative interpretation of (\ref{Taylor}) follows from the observation that the $k$th particle has an electric-dipole moment 
\begin{equation}
\hat{\mathbf{d}}_k=q_k\left(\hat{\bar{\mathbf{r}}}_k'-\hat{\mathbf{r}}_k\right)
\end{equation}
with respect to $\hat{\mathbf{r}}_k$ (but no electric-dipole moment with respect to $\hat{\bar{\mathbf{r}}}_k'$, again in accord with our interpretation of $\hat{\bar{\mathbf{r}}}_k'$ as the particle's centre of charge): the familiar spin-orbit interaction with Thomas precession included \cite{Uhlenbeck26a,Thomas26a,Thomas27a,Gunther-Mohr54a} is also the electric-dipole interaction $\hat{\mathbf{d}}_k\cdot\pmb{\nabla}_k\hat{\Phi}_k^q$. $\hat{\mathbf{d}}_i$ is not to be confused with the more fundamental electron electric-dipole moment currently being sought in high-precision experiments, which is believed to be parallel to the electron's spin, in violation of parity symmetry and time reversal symmetry \cite{Hudson11a, Baron17a, Cairncross17a}. The third line on the right-hand side of (\ref{W0explicit}) can be recast as
\begin{eqnarray}
\frac{1}{2}\sum_k\frac{q_k}{m_k}\hat{\mathbf{p}}_k\cdot\hat{\mathbf{A}}_k&=&\frac{1}{4}\sum_kq_k\left(\hat{\bar{\mathbf{v}}}_k'\cdot\hat{\mathbf{A}}_k+\hat{\mathbf{A}}_k\cdot\hat{\bar{\mathbf{v}}}_k'\right) \nonumber \\
&+&\textrm{O}(\tfrac{1}{c^3}),
\end{eqnarray}
in accord with the idea that $\hat{\bar{\mathbf{v}}}_k'$ is the velocity of the centre of charge of the $k$th particle. The fourth and fifth lines require no further development. Thus,
\begin{eqnarray}
\hat{W}&=&\sum_k m_k c^2+\frac{1}{2}\sum_k\frac{\hat{\pi}_k^2}{m_k}-\frac{1}{8c^2}\sum_k\frac{\hat{\pi}_k^4}{m_k^3} \nonumber \\
&+&\frac{1}{2}\sum_kq_k\hat{\Phi}_k^q+\sum_kq_k(\hat{\bar{\mathbf{r}}}_k'-\hat{\mathbf{r}}_k)\cdot\pmb{\nabla}_k\hat{\Phi}_k^q \nonumber \\
&+&\frac{1}{4}\sum_kq_k\left(\hat{\bar{\mathbf{v}}}_k'\cdot\hat{\mathbf{A}}_k+\hat{\mathbf{A}}_k\cdot\hat{\bar{\mathbf{v}}}_k'\right) \nonumber \\
&+&\frac{1}{6}\sum_kq_kR_k^2\nabla_k^2\hat{\Phi}_k^q-\frac{1}{2}\sum_k \hat{\mathbf{m}}_k\cdot(\pmb{\nabla}_k\times\hat{\mathbf{A}}_k) \nonumber \\
&+&\frac{1}{3}\sum_j\hat{\Theta}_{jab}\partial_{ja}\partial_{jb}\hat{\Phi}_j^q+\textrm{O}(\tfrac{1}{c^3}). \label{moleculeenergy}
\end{eqnarray}
The first line on the right-hand side of (\ref{moleculeenergy}) resembles (\ref{Wmech}), the second and third lines together resemble (\ref{Wfield}) and the remaining lines describe additional contributions due to the finite sizes of the particles, magnetic-dipole moments of the particles and electric-quadrupole moments of the nuclei.


\subsection{Linear momentum}
\label{Linear momentum}
\subsubsection{Classical system}
The total linear momentum of a collection of structureless point particles interacting electromagnetically in the classical domain is \cite{Landau, Griffiths99a, Jackson99a}
\begin{equation}
\tilde{\mathbf{P}}=\tilde{\mathbf{P}}_\textrm{kin}+\tilde{\mathbf{P}}_\textrm{em},
\end{equation}
with
\begin{eqnarray}
\tilde{\mathbf{P}}_\textrm{kin}&=&\sum_k\tilde{\pmb{\pi}}_k \nonumber \\
&=&\sum_k \sqrt{1+\frac{\tilde{\pi}_k^2}{c^2\tilde{m}_k^2}}\tilde{m}_k\tilde{\mathbf{v}}_k \nonumber \\
&=&\sum_k \tilde{m}_k\tilde{\mathbf{v}}_k+\frac{1}{2c^2}\sum_k\frac{\tilde{\pi}_k^2\tilde{\mathbf{v}}_k}{\tilde{m}_k}+\textrm{O}(\tfrac{1}{c^3}) \label{Pmech}
\end{eqnarray}
the total kinetic linear momentum of the particles and
\begin{eqnarray}
\tilde{\mathbf{P}}_\textrm{em}&=&\epsilon_0\int\tilde{\mathbf{E}}\times\tilde{\mathbf{B}}\textrm{d}^3\mathbf{r} \nonumber \\
&=&\int\tilde{\rho}\tilde{\mathbf{A}}\textrm{d}^3\mathbf{r}+\textrm{O}(\tfrac{1}{c^3}) \nonumber \\
&=&\sum_k\tilde{q}_k\tilde{\mathbf{A}}(\tilde{\mathbf{r}}_k) +\textrm{O}(\tfrac{1}{c^3}) \label{Pfield}
\end{eqnarray}
the total linear momentum of the electromagnetic field.

\subsubsection{Molecule}
We identify the total linear momentum of our molecule as being
\begin{equation}
\hat{\mathbf{P}}=\sum_k \hat{\mathbf{p}}_k. \label{Ptotal}
\end{equation}
$\hat{\mathbf{P}}$ generates translations in space:
\begin{eqnarray}
\textrm{e}^{\textrm{i}\mathbf{d}\cdot\hat{\mathbf{P}}/\hbar}\hat{\mathbf{s}}_k\textrm{e}^{-\textrm{i}\mathbf{d}\cdot\hat{\mathbf{P}}/\hbar}&=&\hat{\mathbf{s}}_k, \label{Ptranslate1} \\
\textrm{e}^{\textrm{i}\mathbf{d}\cdot\hat{\mathbf{P}}/\hbar}\hat{\mathbf{r}}_k\textrm{e}^{-\textrm{i}\mathbf{d}\cdot\hat{\mathbf{P}}/\hbar}&=&\hat{\mathbf{r}}_k+\mathbf{d} \\
\textrm{e}^{\textrm{i}\mathbf{d}\cdot\hat{\mathbf{P}}/\hbar}\hat{\mathbf{p}}_k\textrm{e}^{-\textrm{i}\mathbf{d}\cdot\hat{\mathbf{P}}/\hbar}&=&\hat{\mathbf{p}}_k. \label{Ptranslate3}
\end{eqnarray}
The conservation law
\begin{equation}
\frac{\textrm{d}\hat{\mathbf{P}}}{\textrm{d}t}=0 \label{linearmomentumconservation}
\end{equation}
shows that this is a symmetry transformation for the molecule \cite{Noether18a}, embodying invariance under translations in space.

A more explicit expression for $\hat{\mathbf{P}}$ follows from (\ref{Particlecanonicalp}):
\begin{eqnarray}
\hat{\mathbf{P}}&=&\sum_km_k\hat{\bar{\mathbf{v}}}_k+\frac{1}{4c^2}\sum_k\frac{\hat{\pi}_k^2\hat{\bar{\mathbf{v}}}_k+\hat{\bar{\mathbf{v}}}_k\hat{\pi}_k^2}{m_k}\nonumber \\
&+&\sum_k q_k\hat{\mathbf{A}}_k \nonumber \\
&-&\sum_k\frac{\hat{\mathbf{m}}_k\times\pmb{\nabla}_k\hat{\Phi}_k^q}{c^2}+\textrm{O}(\tfrac{1}{c^3}). \label{moleculelinearmomentum}
\end{eqnarray}
The first line on the right-hand side of (\ref{moleculelinearmomentum}) resembles (\ref{Pmech}), the second line resembles (\ref{Pfield}) and the third line is the total hidden momentum of the molecule, which is cancelled by the linear momentum of the intramolecular electromagnetic field \cite{Cameron18a}:
\begin{equation}
-\sum_k \frac{\hat{\mathbf{m}}_k\times \pmb{\nabla}_k\hat{\Phi}_k^q}{c^2}+\sum_k q_k \hat{\mathbf{A}}^\mathbf{m}_k=0, \label{hiddenishidden}
\end{equation}
as one might expect \cite{Thomson04a,Thomson04b,Thomson04c, Griffiths99a, Babson09a}.


\subsection{Angular momentum}
\label{Angular momentum}
\subsubsection{Classical system}
The total angular momentum (about the spatial origin) of a collection of structureless point particles interacting electromagnetically in the classical domain is \cite{Landau, Griffiths99a, Jackson99a}
\begin{equation}
\tilde{\mathbf{J}}=\tilde{\mathbf{J}}_\textrm{kin}+\tilde{\mathbf{J}}_\textrm{em},
\end{equation}
with
\begin{eqnarray}
\tilde{\mathbf{J}}_\textrm{kin}&=&\sum_k\tilde{\mathbf{r}}_k\times\tilde{\pmb{\pi}}_k \nonumber \\
&=&\sum_k\tilde{\mathbf{r}}_k\times\sqrt{1+\frac{\tilde{\pi}_k^2}{c^2\tilde{m}_k^2}}\tilde{m}_k\tilde{\mathbf{v}}_k \nonumber \\
&=&\sum_k\tilde{\mathbf{r}}_k\times\left(\tilde{m}_k\tilde{\mathbf{v}}_k+\frac{\tilde{\pi}_k^2\tilde{\mathbf{v}}_k}{2c^2\tilde{m}_k}\right)+\textrm{O}(\tfrac{1}{c^3}) \label{Jmech}
\end{eqnarray}
the total kinetic angular momentum of the particles and
\begin{eqnarray}
\tilde{\mathbf{J}}_\textrm{em}&=&\epsilon_0\int\mathbf{r}\times\left(\tilde{\mathbf{E}}\times\tilde{\mathbf{B}}\right)\textrm{d}^3\mathbf{r} \nonumber \\
&=&\int\mathbf{r}\times\tilde{\rho}\tilde{\mathbf{A}}\textrm{d}^3\mathbf{r} \nonumber \\
&=&\sum_k\tilde{\mathbf{r}}_k\times\tilde{q}_k\tilde{\mathbf{A}}(\tilde{\mathbf{r}}_k)+\textrm{O}(\tfrac{1}{c^3}) \label{Jfield}
\end{eqnarray}
the total angular momentum of the electromagnetic field.

\subsubsection{Molecule}
We identify the total angular momentum of our molecule as being
\begin{equation}
\hat{\mathbf{J}}=\sum_k\hat{\mathbf{r}}_k\times\hat{\mathbf{p}}_k+\sum_k\hat{\mathbf{s}}_k. \label{Jtotalbland}
\end{equation}
$\hat{\mathbf{J}}$ obeys the usual angular momentum commutation relations:
\begin{eqnarray}
\left[\hat{J}_a,\hat{J}_b\right]=\textrm{i}\hbar\epsilon_{abc}\hat{J}_c,
\end{eqnarray}
and generates rotations in space:
\begin{eqnarray}
\textrm{e}^{\textrm{i}\pmb{\theta}\cdot\hat{\mathbf{J}}/\hbar}\hat{\mathbf{s}}_k\textrm{e}^{-\textrm{i}\pmb{\theta}\cdot\hat{\mathbf{J}}/\hbar}&=&\hat{\mathbf{s}}_k+\pmb{\theta}\times\hat{\mathbf{s}}_k+\textrm{O}(\pmb{\theta}^2),\label{Jrotate1} \\
\textrm{e}^{\textrm{i}\pmb{\theta}\cdot\hat{\mathbf{J}}/\hbar}\hat{\mathbf{r}}_k\textrm{e}^{-\textrm{i}\pmb{\theta}\cdot\hat{\mathbf{J}}/\hbar}&=&\hat{\mathbf{r}}_k+\pmb{\theta}\times\hat{\mathbf{r}}_k+\textrm{O}(\pmb{\theta}^2) \\
\textrm{e}^{\textrm{i}\pmb{\theta}\cdot\hat{\mathbf{J}}/\hbar}\hat{\mathbf{p}}_k\textrm{e}^{-\textrm{i}\pmb{\theta}\cdot\hat{\mathbf{J}}/\hbar}&=&\hat{\mathbf{p}}_k+\pmb{\theta}\times\hat{\mathbf{p}}_k+\textrm{O}(\pmb{\theta}^2). \label{Jrotate3}
\end{eqnarray}
The conservation law
\begin{equation}
\frac{\textrm{d}\hat{\mathbf{J}}}{\textrm{d}t}=0 \label{Jconservation}
\end{equation}
shows that this is a symmetry transformation for the molecule \cite{Noether18a}, embodying invariance under rotations in space.

A more explicit expression for $\hat{\mathbf{J}}$ follows from (\ref{Particlecanonicalp}):
\begin{eqnarray}
\hat{\mathbf{J}}&=&\sum_k\mathbf{r}_k\times\left(m_k\hat{\bar{\mathbf{v}}}_k+\frac{\hat{\pi}_k^2\hat{\bar{\mathbf{v}}}_k+\hat{\bar{\mathbf{v}}}_k\hat{\pi}_k^2}{4c^2m_k}\right) \nonumber \\
&+&\sum_k\mathbf{r}_k\times q_k\hat{\mathbf{A}}_k \nonumber \\
&+&\sum_k\mathbf{r}_k\times\left(-\frac{\hat{\mathbf{m}}_k\times\pmb{\nabla}_k\hat{\Phi}_k^q}{c^2}\right) \nonumber  \\
&+&\sum_k\hat{\mathbf{s}}_k+\textrm{O}(\tfrac{1}{c^3}). \label{moleculerotationangularmomentum}
\end{eqnarray}
The first line on the right-hand side of (\ref{moleculerotationangularmomentum}) resembles (\ref{Jmech}), the second line resembles (\ref{Jfield}), the third line is the total hidden angular momentum of the molecule and the fourth line is the total mean spin of the molecule. It seems that the total hidden angular momentum of the molecule is \textit{not} cancelled by the angular momentum of the intramolecular electromagnetic field: we note in particular that
\begin{equation}
\sum_k\mathbf{r}_k\times\left(-\frac{\hat{\mathbf{m}}_k\times\pmb{\nabla}_k\hat{\Phi}_k^q}{c^2}+q_k \hat{\mathbf{A}}^\mathbf{m}_k\right)\ne 0,
\end{equation}
in spite of the fact that the total hidden (\textit{linear}) momentum of the molecule is cancelled by the linear momentum of the intramolecular electromagnetic field, as shown in (\ref{hiddenishidden}). This represents no fundamental difficulty, as the total angular momentum of a system `at rest' need not vanish \cite{Griffiths99a}.


\subsection{Boost momentum}
\label{Boost momentum}
The total energy of a system generates translations of the system in time, the total linear momentum generates translations in space and the total angular momentum generates circular rotations in space. Similarly, the total boost momentum of a system generates hyperbolic rotations of the system in spacetime, also known as Lorentz transformations or boosts \cite{Barnett11a, Cameron12a}. The conservation of boost momentum embodies invariance under such rotations \cite{Bessel-Hagen21a, Cameron12a, Bliokh13a}. When taken together with the global conservation of energy and linear momentum, the global conservation of boost momentum can be interpreted as a statement that the system translates with constant velocity \cite{Landau, Bessel-Hagen21a, Bliokh12a}.

\subsubsection{Classical system}
The total boost momentum (about the spacetime origin) of a collection of structureless point particles interacting electromagnetically in the classical domain is \cite{Landau}
\begin{equation}
\tilde{\mathbf{K}}=t\tilde{\mathbf{P}}-\tilde{\mathbf{D}}_\textrm{kin}-\tilde{\mathbf{D}}_\textrm{em},
\end{equation}
with
\begin{eqnarray}
\tilde{\mathbf{D}}_\textrm{kin}&=&\sum_k\sqrt{1+\frac{\tilde{\pi}_k^2}{c^2\tilde{m}_k^2}}\tilde{m}_k\tilde{\mathbf{r}}_k \nonumber \\
&=&\sum_k\tilde{m}_k\tilde{\mathbf{r}}_k+\frac{1}{2c^2}\sum_k\frac{\tilde{\pi}_k^2\tilde{\mathbf{r}}_k}{\tilde{m}_k} \nonumber \\
&+&\textrm{O}(\tfrac{1}{c^3}) \label{Kmech}
\end{eqnarray}
defining the total kinetic boost momentum $t\tilde{\mathbf{P}}_\textrm{kin}-\tilde{\mathbf{D}}_\textrm{kin}$ of the particles and
\begin{eqnarray}
\tilde{\mathbf{D}}_\textrm{em}&=&\frac{\epsilon_0}{2c^2}\int \mathbf{r}\left(\tilde{E}^2+c^2\tilde{B}^2\right)\textrm{d}^3\mathbf{r} \nonumber \\
&=&\frac{1}{2c^2}\int\mathbf{r}\tilde{\rho}\tilde{\Phi}\textrm{d}^3\mathbf{r}+\textrm{O}(\tfrac{1}{c^3}) \nonumber \\
&=&\frac{1}{2c^2}\sum_k\tilde{\mathbf{r}}_k\tilde{q}_k\tilde{\Phi}(\tilde{\mathbf{r}}_k)+\textrm{O}(\tfrac{1}{c^3}) \label{Kfield}
\end{eqnarray}
defining the total boost momentum $t\tilde{\mathbf{P}}_\textrm{em}-\tilde{\mathbf{D}}_\textrm{em}$ of the electromagnetic field.

\subsubsection{Molecule}
It seems that there is no closed form for the total boost momentum of our molecule. This is unsurprising, as the molecular Hamiltonian $\hat{H}$ is not Lorentz invariant but rather is effectively truncated at order $1/c^2$. Working to order $1/c^2$ then, we identify the total boost momentum of the molecule as being
\begin{equation}
\hat{\mathbf{K}}=t\hat{\mathbf{P}}-\hat{\mathbf{D}} \label{Kusualform}
\end{equation}
with
\begin{eqnarray}
\hat{\mathbf{D}}&=&\sum_k m_k\hat{\bar{\mathbf{r}}}_k+\frac{1}{4c^2}\sum_k\frac{\hat{\pi}_k^2\hat{\bar{\mathbf{r}}}_k+\hat{\bar{\mathbf{r}}}_k\hat{\pi}_k^2}{m_k}\nonumber \\
&+&\frac{1}{2c^2}\sum_k \hat{\bar{\mathbf{r}}}_kq_k\hat{\Phi}_k^q+\textrm{O}(\tfrac{1}{c^3}), \label{moleculeD}
\end{eqnarray}
which suffices to ensure the conservation law
\begin{eqnarray}
\frac{\textrm{d}\hat{\mathbf{K}}}{\textrm{d}t}&=&\frac{\textrm{i}}{\hbar}\left[\hat{H},\hat{\mathbf{K}}\right]+\frac{\partial}{\partial t}\hat{\mathbf{K}} \nonumber \\
&=&0+\textrm{O}(\tfrac{1}{c^3}), \label{explicitboostconservation}
\end{eqnarray}
as desired. The first line on the right-hand side of (\ref{moleculeD}) resembles (\ref{Kmech}) and the second line resembles (\ref{Kfield}). Note that it is necessary to identify the kinetic position $\hat{\bar{\mathbf{\mathbf{r}}}}_k$ in the first term, rather than the mean position $\hat{\mathbf{r}}_k$, for example.

In general, the rate of change with respect to time of an energy is referred to as a power, the rate of change of a linear momentum is referred to as a force and the rate of change of an angular momentum is referred to as a torque. There does not appear to be a unique terminology, however, for the rate of change of a boost momentum \cite{Bessel-Hagen21a, Landau, Barnett11a, Bliokh12a, Cameron12a, Bliokh13a, Bliokh18a}, which is surprising given the fundamental importance of boost momentum. We propose, therefore, that the rate of change of a boost momentum be referred to henceforth as a `jig'. The need to distinguish between a torque and a jig is \textit{not} negated by the fact that angular momentum and boost momentum reside together in a tensor, any more than the need to distinguish between a power and a force is negated by the fact that energy and linear momentum reside together in a tensor \cite{Bessel-Hagen21a, Landau, Barnett11a, Bliokh12a, Cameron12a, Bliokh13a}.


\section{Outlook}
\label{Outlook}
Possible avenues for future research into the relativistic properties of a molecule include the following.
\begin{itemize}
\item The identification of generalised translational, rotational, vibrational and electronic coordinates to order $1/c^2$.
\item The separation of angular momentum into spin and orbital or intrinsic and extrinsic parts to order $1/c^2$.
\item The interaction of a molecule with externally imposed fields to order $1/c^2$. In particular, the calculation of powers, forces, torques and jigs, which can change the energy, linear momentum, angular momentum and boost momentum of the molecule.
\end{itemize} We will return to these and related tasks elsewhere.


\section{Acknowledgements}
This work was supported by the EPSRC (EP/M004694/1) and The Leverhulme Trust (RPG-2017-048). We thank Gergely Ferenczi and an anonymous referee for their useful comments and suggestions.


\begin{appendix}
\section{Breakdown of $\hat{H}$}
The molecular Hamiltonian can be partitioned in the form
\begin{equation}
\hat{H}=\hat{H}_\textrm{basic}+\hat{H}_\textrm{fs}+\hat{H}_\textrm{hfs},
\end{equation}
with $\hat{H}_\textrm{basic}$, $\hat{H}_\textrm{fs}$ and $\hat{H}_\textrm{hfs}$ as described below.

The basic molecular Hamiltonian
\begin{eqnarray}
\hat{H}_\textrm{basic}&=&c^2\sum_k m_k+\frac{1}{2}\sum_i\frac{\hat{p}_i^2}{m_i}+\frac{1}{8\pi\epsilon_0}\sum_k\sum_{k'\ne k}\frac{ q_k q_{k'}}{\hat{r}_{kk'}} \nonumber\\
&+&\frac{1}{2}\sum_j\frac{\hat{p}^2_j}{m_j} 
\label{H0}
\end{eqnarray}
accounts for the basic structure of the molecule. In a notional order of decreasing `size'...
\begin{itemize}
\item The first term on the right-hand side of (\ref{H0}) is the total rest energy of the particles. 
\item The second term describes the kinetic energies of the electrons to order $1/c^0$.
\item The third term describes the Coulomb interactions between the particles.
\item The fourth term describes the kinetic energies of the nuclei to order $1/c^0$.
\end{itemize}
The first term is constant and so is usually omitted.

The fine-structure Hamiltonian
\begin{eqnarray}
\hat{H}_\textrm{fs}&=&-\frac{1}{8 c^2}\sum_i \frac{\hat{p}_i^4}{m_i^3} \nonumber \\
&-&\frac{\hbar^2}{8\epsilon_0 c^2}\sum_i \frac{q_i}{m_i^2} \sum_jq_j\delta^3(\hat{\mathbf{r}}_{ij}) \nonumber \\
&+&\frac{1}{4\pi \epsilon_0 c^2} \sum_{i} \left(1-\frac{q_i}{2 m_i \gamma_i}\right) \hat{\mathbf{m}}_i\cdot\left(\frac{\hat{\mathbf{p}}_i}{m_i}\times\sum_{k\ne i}\frac{q_k   \hat{\mathbf{r}}_{ik}}{\hat{r}_{ik}^3}\right) \nonumber \\
&+&\frac{\mu_0}{4\pi} \sum_{i}\hat{\mathbf{m}}_i\cdot\sum_{i'\ne i}\frac{q_{i'} \hat{\mathbf{r}}_{ii'}\times \hat{\mathbf{p}}_{i'}}{m_{i'} \hat{r}_{ii'}^{3}} \nonumber \\
&+&\frac{\mu_0}{8\pi}\sum_i\sum_{i' \ne i}\Bigg\{\frac{1}{\hat{r}_{ii'}^3}\Bigg[ \hat{\mathbf{m}}_i\cdot\hat{\mathbf{m}}_{i'}-\frac{3 (\hat{\mathbf{m}}_i\cdot\hat{\mathbf{r}}_{ii'}) (\hat{\mathbf{m}}_{i'}\cdot\hat{\mathbf{r}}_{ii'})}{\hat{r}_{ii'}^2} \Bigg] \nonumber \\
&-&\frac{8\pi}{3}\hat{\mathbf{m}}_i \cdot\hat{\mathbf{m}}_{i'}\delta^3(\hat{\mathbf{r}}_{ii'})  \Bigg\} \nonumber \\
&-&\frac{\hbar^2}{8\epsilon_0 c^2}\sum_i \frac{q_i}{m_i^2} \sum_{i'\ne i}q_{i'}\delta^3(\hat{\mathbf{r}}_{ii'}) \nonumber \\
&-&\frac{\mu_0}{16\pi}\sum_i\sum_{i'\ne i} \frac{q_i q_{i'}}{m_i m_{i'}}\Bigg[\hat{p}_{ia} \frac{1}{ \hat{r}_{ii'} } \hat{p}_{i'a} \nonumber \\
&+&\left(\hat{\mathbf{p}}_i\cdot\hat{\mathbf{r}}_{ii'}\right)\frac{1}{\hat{r}_{ii'}^3}\left(\hat{\mathbf{r}}_{ii'}\cdot\hat{\mathbf{p}}_{i'}\right)\Bigg] \nonumber \\
&+&\frac{\mu_0}{4\pi} \sum_{i}\hat{\mathbf{m}}_i\cdot\sum_j\frac{q_{j} \hat{\mathbf{r}}_{ij}\times \hat{\mathbf{p}}_j}{m_j\hat{r}_{ij}^{3}} \nonumber \\
&-&\frac{\mu_0}{16\pi}\sum_j\sum_i \frac{q_j q_i}{m_j m_i}\Bigg[\hat{p}_{ja} \frac{1}{ \hat{r}_{ji}} \hat{p}_{ia}+\hat{p}_{ia}\frac{1}{\hat{r}_{ji}} \hat{p}_{ja} \nonumber \\
&+&\left(\hat{\mathbf{p}}_j\cdot\hat{\mathbf{r}}_{ji}\right)\frac{1}{\hat{r}_{ji}^3}\left(\hat{\mathbf{r}}_{ji}\cdot\hat{\mathbf{p}}_i\right)+\left(\hat{\mathbf{p}}_i\cdot\hat{\mathbf{r}}_{ji}\right)\frac{1}{\hat{r}_{ji}^3}\left(\hat{\mathbf{r}}_{ji}\cdot\hat{\mathbf{p}}_j\right)\Bigg] \nonumber \\
&-&\frac{\mu_0}{16\pi}\sum_j\sum_{j'\ne j} \frac{q_j q_{j'}}{m_j m_{j'}}\Bigg[\hat{p}_{ja} \frac{1}{\hat{r}_{jj'}} \hat{p}_{j'a} \nonumber \\
&+&\left(\hat{\mathbf{p}}_j\cdot\hat{\mathbf{r}}_{jj'}\right)\frac{1}{\hat{r}_{jj'}^3}\left(\hat{\mathbf{r}}_{jj'}\cdot\hat{\mathbf{p}}_{j'}\right)\Bigg] \nonumber \\
&-&\frac{1}{8 c^2}\sum_j \frac{\hat{p}_j^4}{m_j^3} 
\label{middle}
\end{eqnarray}
accounts for certain higher-order intramolecular interactions in which the $\hat{m}_j$, $\hat{\Theta}_{jab}$ and $R_j$ do not appear. In a notional order of decreasing `size'...
\begin{itemize}
\item The first term on the right-hand side of (\ref{middle}) describes relativistic corrections to the kinetic energies of the electrons \cite{Gaunt29a, Breit29a}. 
\item The second term describes one-electron Darwin interactions \cite{Darwin28a}.
\item The third term describes electron spin-orbit interactions \cite{Uhlenbeck26a}, with Thomas precession \cite{Thomas26a,Thomas27a} included.
\item The fourth term describes electron-spin / other-electron-orbit interactions \cite{Heisenberg26a}.
\item The fifth term describes electron-spin / electron-spin interactions, comprised of classical dipolar \cite{Heisenberg26a} and contact \cite{Gaunt29a} contributions.
\item The sixth term describes two-electron Darwin interactions \cite{Gaunt29a, Breit29a}.
\item The seventh term describes electron-orbit / electron-orbit interactions, comprised of magnetic \cite{Gaunt29a} and retardation \cite{Breit29a} contributions.
\item The eighth term describes electron-spin / nuclear-orbit interactions.
\item The ninth term describes electron-orbit / nuclear-orbit interactions, comprised of magnetic and retardation contributions.
\item The tenth term describes nuclear-orbit / nuclear-orbit interactions, comprised of magnetic and retardation contributions.
\item The eleventh term describes relativistic corrections to the kinetic energies of the nuclei.
\end{itemize}
The ninth, tenth and eleventh terms are particularly `small' and are often neglected.

The hyperfine-structure Hamiltonian
\begin{eqnarray}
\hat{H}_\textrm{hfs}&=&\frac{\mu_0}{4\pi} \sum_j\hat{\mathbf{m}}_j\cdot\sum_i\frac{q_i\hat{\mathbf{r}}_{ji}\times \hat{\mathbf{p}}_i}{m_i\hat{r}_{ji}^{3}}  \nonumber \\
&+&\frac{\mu_0}{4\pi}\sum_i\sum_j\Bigg\{\frac{1}{\hat{r}_{ij}^3}\Bigg[ \hat{\mathbf{m}}_i\cdot\hat{\mathbf{m}}_j-\frac{3 (\hat{\mathbf{m}}_i\cdot\hat{\mathbf{r}}_{ij}) (\hat{\mathbf{m}}_j\cdot\hat{\mathbf{r}}_{ij})}{\hat{r}_{ij}^2} \Bigg] \nonumber \\
&-&\frac{8\pi}{3}\hat{\mathbf{m}}_i \cdot\hat{\mathbf{m}}_j\delta^3(\hat{\mathbf{r}}_{ij}) \Bigg\} \nonumber \\
&-&\frac{1}{12\pi \epsilon_0}\sum_j \hat{\Theta}_{jab}  \sum_{k\ne j} q_k \left(\frac{\delta_{ab}}{\hat{r}_{jk}^3}-\frac{3\hat{r}_{jk a}\hat{r}_{jk b}}{\hat{r}_{jk}^5}\right) \nonumber \\
&-&\frac{1}{6\epsilon_0}\sum_iq_i\sum_j q_j R_j^2\delta^3(\hat{\mathbf{r}}_{ij}) \nonumber \\
&+&\frac{1}{4\pi \epsilon_0 c^2} \sum_j\left(1-\frac{q_j}{2m_j\gamma_j}\right)\hat{\mathbf{m}}_j\cdot\left(\frac{\hat{\mathbf{p}}_j}{m_j}\times\sum_{k\ne j}\frac{q_k\hat{\mathbf{r}}_{jk}}{\hat{r}_{jk}^3}\right) \nonumber \\
&+&\frac{\mu_0}{4\pi} \sum_j\hat{\mathbf{m}}_j\cdot\sum_{j'\ne j}\frac{q_{j'}\hat{\mathbf{r}}_{jj'}\times \hat{\mathbf{p}}_{j'}}{m_{j'}\hat{r}_{jj'}^{3}} \nonumber \\
&+&\frac{\mu_0}{8\pi}\sum_j\sum_{j'\ne j}\frac{1}{\hat{r}_{jj'}^3}\Bigg[\hat{\mathbf{m}}_j\cdot\hat{\mathbf{m}}_{j'}-\frac{3 (\hat{\mathbf{m}}_j\cdot\hat{\mathbf{r}}_{jj'}) (\hat{\mathbf{m}}_{j'}\cdot\hat{\mathbf{r}}_{jj'})}{\hat{r}_{jj'}^2} \nonumber \\
&-&\frac{8\pi}{3}\hat{\mathbf{m}}_i \cdot\hat{\mathbf{m}}_j\delta^3(\hat{\mathbf{r}}_{ij})\Bigg] \nonumber \\
&-&\frac{1}{6\epsilon_0}\sum_jq_j\sum_{j'\ne j} q_{j'} R_{j'}^2\delta^3(\hat{\mathbf{r}}_{jj'}) \label{Hhfs}
\end{eqnarray}
accounts for certain higher-order intramolecular interactions in which the $\hat{m}_j$, $\hat{\Theta}_{jab}$ and $R_j$ appear. In a notional order of decreasing `size'...
\begin{itemize}
\item The first term on the right-hand side of (\ref{Hhfs}) describes nuclear-spin / electron-orbit interactions \cite{Henderson48a}.
\item The second term describes electron-spin / nuclear-spin interactions, comprised of classical dipolar and Fermi contact \cite{Fermi30a} contributions. 
\item The third term describes nuclear electric-quadrupole interactions \cite{Casimir36a}.
\item The fourth term describes nuclear finite-size corrections to the electron-nucleus Coulomb interactions \cite{Helgaker}.
\item The fifth term describes nuclear spin-orbit interactions, with Thomas precession included \cite{Gunther-Mohr54a}.
\item The sixth term describes nuclear-spin / other-nucleus-orbit interactions \cite{Gunther-Mohr54a}.
\item The seventh term describes nuclear-spin / nuclear-spin interactions, comprised of classical dipolar \cite{Kellogg39a} and contact contributions.
\item The eighth term describes nuclear finite-size corrections to the nucleus-nucleus Coulomb interactions.
\end{itemize}
The contact contributions to the seventh term, as well as the eighth term, are particularly `small' and are often neglected.

Note that the nuclei are effectively described in the basic molecular Hamiltonian $\hat{H}_\textrm{basic}$ and the fine structure Hamiltonian $\hat{H}_\textrm{fs}$ as structureless point particles with electric charge: we have attributed interactions that involve nuclear magnetic-dipole moments, nuclear electric-quadrupole moments or nuclear finite-size corrections to the hyperfine structure Hamiltonian $\hat{H}_\textrm{hfs}$. Our reasoning is that interactions that involve nuclear magnetic-dipole moments or nuclear electric-quadrupole moments are nuclear-spin dependent and nuclear finite-size corrections emerge at the same order of multipole expansion. One might argue moreover that nuclear \textit{Zitterbewegung} is nuclear-spin dependent \cite{Pachucki95a}.

\end{appendix}



\bibliographystyle{apsrev4-1}
\bibliography{WPJK}


\end{document}